\documentclass[useAMS,usenatbib,usegraphicx,referee]{mn2e}
\usepackage{amssymb}

\title[Could AX J1841.0$-$0536 Be an Anti-Magnetar?]
{Could AX J1841.0$-$0536 Be an Anti-Magnetar?}

\author[Li \& Zhang]
{X.-D. Li\thanks{E-mail: lixd@nju.edu.cn} and Z. Zhang\\
Department of Astronomy and Key Laboratory of Modern Astronomy and Astrophysics,
Nanjing University, Nanjing 210093, China
 }

\begin{document}
\date{Accepted  . Received  ; in original form }

\pagerange{\pageref{firstpage}--\pageref{lastpage}} \pubyear{2002}

\maketitle

\begin{abstract}
Recent observations show that  supergiant fast X-ray transients (SFXTs) 
spend most of their lifetime at an intermediate level luminosity 
$\sim 10^{33}-10^{34}$ ergs$^{-1}$, and,  when a blackbody model for the spectra is adopted, 
the resulting radii of the emission region are always only a few hundred meters,
supporting the idea that during the intermediate state SFXTs are 
accreting matter from the companion star. From these observational phenomena
we derive possible constraints on the magnetic field strengths of the 
neutron stars in four SFXTs with known spin periods. While IGR J11215$-$5952, 
IGR J16465$-$4507, and IGR J18483$-$0311 may have 
magnetic fields (up to a few $10^{11}-10^{12}$ G) similar to those of normal
X-ray pulsars, 
the magnetic field of AX J1841.0$-$0536 is considerably low ($\la 10^{10}$ G). 
The high-mass companion stars in SFXTs implies that the neutron stars
are relatively young objects, with age less than $\sim 10^7$ yr.
Analysis of the spin evolution shows that neutron stars like 
AX J1841.0$-$0536 should be 
born with relatively long spin periods ($\la 1$ s). 
Considering the fact that among the four SFXTs only AX J1841.0$-$0536 is a ``proper" one, and
the other three are either ``intermediate" SFXTs or have peculiar characteristics,
we suggest that the neutron stars in some of SFXTs may have
similar characteristics as several young central compact objects in 
supernova remnants called ``anti-magnetars". These features, combining
with accretion from clumpy winds, could 
make them distinct from standard supergiant X-ray binaries -- the low fields and
relatively long spin periods guarantee accretion at very low level, resulting in a
large dynamic range ($10^4-10^5$) of X-ray luminosity. 
\end{abstract}

\begin{keywords}
stars: neutron -- X-rays: binaries -- X-rays: individuals: IGR J11215$-$5952, 
IGR J16465$-$4507, IGR J18483$-$0311,  AX J1841.0$-$0536
\end{keywords}

\section{Introduction}
Most of the high-mass X-ray binaries (HMXBs) consist of a neutron
star (NS)  accreting from its high-mass companion star, usually of O
or B type. They are traditionally divided into two types, in which
either a supergiant star or a Be star is contained, and the X-ray
source often is a pulsar. In the supergiant systems, accretion
occurs via stellar wind capture and beginning atmosphere Roche-lobe
overflow, while in the Be systems only wind accretion takes place,
since the Be star is well inside its Roche-lobe \citep[see][for a review]{Tau06}. In
recent years, a new type of HMXBs were discovered by {\em INTEGRAL},
called supergiant fast X-ray transients \citep[SFXTs. See e.g.,][for a 
recent review]{Sid11}. These sources are
characterized by X-ray outbursts composed of short bright flares up to 
peak luminosities of $\sim 10^{36}-10^{37}$ ergs$^{-1}$, with duration of a 
few hours for each single flare. In quiescence the luminosity is as low as
$10^{32}$ ergs$^{-1}$, so that the dynamic range of X-ray luminosity
reaches $4-5$ orders of magnitude. X-ray pulsations have been
detected in a few sources, leading to the firm identiÞcation of the compact 
object as an NS. This is also
supported by their hard X-ray spectra, which are represented by flat
hard power law below 10 keV with high energy cut-off at about
$15-30$ keV, sometimes strongly absorbed at soft energies, similar as
in X-ray pulsars
\citep{Wal06,Sid06,RS10}.  

Models for SFXTs can be roughly divided into three categories,
though none of them  can be responsible for all the observational
properties. The most widely studied one is the clumpy wind model,
which invokes the idea that the supergiant winds are strongly
inhomogeneous rather than smooth \citep{Os07}, with large
density contrasts ($\sim 10^4-10^5$). The short flares in SFXTs are
suggested to be produced by accretion of massive clumps
($10^{22}-10^{23}$ g) in the winds \citep{int05,WZ07,Neg08}. 
Between the clumps the
accretion rate is very low, and the sources become dim. The clumps
are expected to be small and crowded near the surface of the supergiant, and become
large and sparse when the wind blows out. As a result, in the region
within $2R_*$ (where $R_*$ is the radius of the supergiant), the
clumpy wind can be treated as roughly homogeneous and continuous,
while outside the region, the effect of clumping becomes important
\citep{Neg08}. In this model, the SFXTs should display wider and 
more eccentric orbits than persistent HMXBs. 

The regular 329 day outbursts and longer outburst duration, as well
as the  narrow and step light curve in IGR J11215$-$5952 suggests the
presence of an inclined equatorial wind component in this system,
which is denser and slower than the symmetric polar wind from the
blue supergiant  \citep{Sid07}. This model requires the outbursts to
be periodical or semi-periodical. Besides IGR J11215$-$5952, 
IGR J18483$-$0311 also displays periodically recurrent outbursts
\citep{Sid09b}, suggesting that the outburst is triggered near the periastron 
passage in a highly eccentric orbit.

The gated models \citep{GS07,Boz08} are based on the analysis of
different stage of the NS's  spin evolution in the wind environment by
comparing the accretion radius $R_a$, the magnetospheric radius
$R_m$  and the corotation radius $R_{co}$.
Transitions across these stages may lead to large variations in the X-ray
luminosity with relatively small changes in the mass accretion rate. 
If some of the SFXTs have long spin periods ($>1000$ s), this model  implies
 large magnetic fields ($\ga 10^{13}-10^{15}$ G) of the NSs in the cases of 
 the centrifugal barrier and the magnetic barrier.

Recent observations with {\em Swift}, {\em Suzaku}, and {\em XMM-Newton}
have revealed some new features of SFXTs 
\citep[e.g.][]{Sid08,Sid09a,Sid10,Giu09,Rom08,Rom09a,Rom09b,Rom10,Rom11,
Boz10,Bod10}. 
One of the most prominent results
is that SFXTs spend most of their life still accreting matter even
outside bright flaring activity rather in quiescence, emitting at an
intermediate level luminosity $\sim 10^{33}-10^{34}$ ergs$^{-1}$ with  
hard X-ray spectra. Especially,
spectral analyses of the X-ray emission in the intermediate state
show that, when a blackbody model is adopted, the resulting radii of
the emission region are always only a few hundred meters, clearly
associated with the polar caps of the NSs. These results
strongly supports the fact that the intermediate (and probably very low)
intensity emission of SFXTs is produced by accretion of matter
onto the NSs. In the following section we use these facts to constrain
the magnetic field strengths of several NSs in SFXTs. We discuss their possible
implications in section 3 and conclude in section 4.

%Though SFXTs exhibit diverse behaviours , like IGR J11215-5952 whose outbursts last 5 days and AXT 1749.1-2733 whose lightcurve profile of the outburst is smooth sufficiently, Swift confirmed their common X-ray characteristics, such as outburst lengths well in excess of hours, with multiple peaked structure, and a high dynamic range (including bright outbursts), up to $\thicksim{4}$ orders of magnitude \citep{Rom10a}.

\section{Possible constraints on NS magnetic field strengths}

\citet{DP81} argued that, when captured by an
NS in HMXBs, the wind material from the optical companion star
always forms a quasi-static envelope surrounding the NS.  The interaction 
between the NS and the wind material leads to different evolutionary 
stages of the NS, i.e., (a) pulsar phase, (b) rapid rotator phase, 
(c) supersonic propeller phase and (d) subsonic propeller phase. 
It is not clear whether phase (d) exists in realistic 
situation (see also discussion below). Due to
the short duration and weak radiation, it is difficult to be testified by either observations or
numerical simulations. Population synthesis calculations by \citet{DL06}
showed that the relation between the orbital periods and the spin periods of wind-fed 
X-ray pulsars in supergiant HMXBs can be roughly accounted without
requiring that an X-ray pulsar emerges when passing phase (d).
Nevertheless, the end of phase (c) can be taken as a  
conservative condition for the occurrence of accretion onto the polar cap region of the NS.
The observational evidence of accretion onto the NS in SFXTs during
the intermediate state may help draw useful constraints on 
the magnetic fields of the NSs. In the following we use $M$, $\dot{M}$, $B$, $R$, and 
$P_{\rm s}$ to denote the mass, mass
accretion rate,  magnetic field, radius, and spin period of the NS, 
respectively

(1) If steady accretion occurs during the intermediate state, one possible
implication is that the plasma at the base of the NS magnetosphere has become
sufficiently cool, so that the magnetospheric boundary is unstable with 
respect to interchange instabilities \citep[e.g. Rayleigh-Taylor instability, or RTI.][]{AL76,EL77}. 
This can be realized only if  the cooling processes in the envelope are
effective and the spin period of the star exceeds a so-called break
period $P_{\rm br}$ \citep{DP81,Ik01a}, which denotes the end of phase (d),
\begin{equation}
P_{\rm s}\ga P_{\rm br}\simeq 1.2\times 10^4B_{12}^{16/21}\dot{M}_{13}^{-5/7}
M_1^{-4/21}R_6^{16/7}\,{\rm s},
\end{equation}
%or equivalently
%\begin{equation}
%B_{12, max}\simeq 2.24\times 10^{-4} (\frac{P_{\rm s}}{20\,{\rm
%s}})^{21/16}\dot{M}_{13}^{15/16}M_1^{1/6}R_6^{-3},
%\end{equation}
where $B_{12}=B/10^{12}$ G, $R_6=R/10^6$ cm, $\dot{M}_{13}=\dot{M}/10^{13}$
gs$^{-1}$,  and $M_1=M/1M_{\odot}$,
\citet{AL76} and \citet{EL77} also showed that
for the conditions of interest the magnetospheric boundary can be interchange 
unstable only if the Compton cooling is effective. This gives the following
condition on the X-ray luminosity,
\begin{equation}
L_{\rm X}\ga L_{\rm cr}\simeq 3\times 10^{36} B_{12}^{1/4}M_1^{1/2}R_6^{-1/8}
{\rm ergs}^{-1}.
\end{equation}
Obviously the luminosity given by Eq.~(2) is too high for the intermediate state luminosities
($\sim 10^{33}-10^{34}$  ergs$^{-1}$) of SFXTs, unless the magnetic fields are extremely 
(and unrealistically) low. Thus we are led to the conclusion  that the X-ray emission
during the intermediate state is unlikely due to accretion driven by
interchange instabilities.

The calculations by \citet{AL76} and \citet{EL77} were all performed 
by assuming a non-rotating NS. This is a reasonable approximation in the case of 
very slowly rotating objects.  \citet{Burn83}  showed that different conditions might 
apply if the rotation of the NS is taken into account. In this case, even if the 
RTI is suppressed, the accreting matter can still penetrate through 
the star's magnetic field lines due to a shearing instability (Kelvin-Helmholtz instability, or KHI).
This means that a relatively large amount of matter can accrete onto the NS 
even if the RTI is not at work. Thus we first take the simplest assumption that direct accretion 
occurs with the help of KHI if the magnetospheric radius of the  NS
\begin{equation}
R_{\rm m}=(\frac{B^2R^6}{\dot{M}\sqrt{2GM}})^{2/7}
\end{equation}
is smaller than the corotation radius
\begin{equation}
R_{\rm co}=(\frac{GM}{4\pi^2}P_{\rm s}^2)^{1/3},
\end{equation}
where $G$ is the gravitational constant. 
Equivalently the NS has evolved beyond phase (c) and its spin period is 
longer than the equilibrium period,
\begin{equation}
P\ga P_{\rm eq}\simeq165.5B_{12}^{6/7}\dot{M}_{13}^{-3/7}M_1^{-5/7}R_6^{18/7}\,{\rm s}.
\end{equation}
This condition sets the maximum of the magnetic field strength of the NS as,
\begin{equation}
B_{12, max}\simeq 0.085(\frac{P_{\rm s}}{20\,{\rm
s}})^{7/6}\dot{M}_{13}^{1/2}M_1^{5/6}R_6^{-3}.
\end{equation}

(2) The above derivations are based on the assumption that there is direct accretion 
during  the intermediate state. This may apply for, e.g., the
clumpy wind model.  In the gated model, however, even if the
magnetospheric boundary is stable against RTI or KHI,  in the subsonic propeller phase (d),
part of the plasma at the base of the NS magnetosphere
may penetrate into the magnetosphere and flow along the field lines to the 
polar caps due to turbulent diffusion and  reconnection of the magnetic field
lines, as suggested by \citet{Ik01b} and \citet{Boz08}.  \citet{Ik01b}  showed that 
the rate of plasma penetration 
due to magnetic line reconnection is about
$1\%$ of the mass capture rate $\dot{M}_{\rm c}$, when the efficiency of
the reconnection process $\alpha_r\sim 0.1$, and the average scale
of plasma vortices of the embedded field is $\lambda_{m}\sim
0.1R_{m}$. The rate of diffusion is generally lower than that of line reconnection by more
than two orders of magnitude, and is not considered here. 
Roughly speaking, to produce X-ray emission with luminosity of $10^{33}-10^{34}$
ergs$^{-1}$ in phase (d), the mass capture rate has to be $\dot{M}_{\rm c}\sim 10^{15}-10^{16}$
gs$^{-1}$. However, the big uncertainties in the
introduced parameters ($\alpha_r$ and $\lambda_{m}$) prevent a 
reliable estimate of the mass capture rates from observed luminosities.
%The condition
%\begin{equation}
%P_{\rm eq}=23\mu_{30}^{6/7}\dot{M}_{c,15}^{-3/7}M_1^{-5/7}\,{\rm s}<P_{\rm s}<P_{\rm br}=
%450\mu_{30}^{16/21}\dot{M}_{c,15}^{-5/7}M_1^{-4/21}\,{\rm s},
%\end{equation}
%where $\dot{M}_{c,15}=\dot{M}_c/10^{15}$ gs$^{-1}$, results in
%\begin{equation}
%1.68\times 10^{-2} (\frac{P_{\rm s}}{20\,{\rm s}})^{21/16}\dot{M}_{c,15}^{15/16}M_1^{1/6}
%<\mu_{30}<0.85(\frac{P_{\rm s}}{20\,{\rm s}})^{7/6}\dot{M}_{c,15}^{1/2}M_1^{5/6}.
%\end{equation}
On the other hand, 
%flares down to $\la 10^{35}$ ergs$^{-1}$ have
%been  confidently identified \citep{Rom10}. 
If the intermediate level X-ray luminosities are really produced by field penetration 
in phase (d), then  we can safely derive that, during outbursts direct accretion must 
take place, i.e., the NS spin period must be longer than the corresponding break period. 
Taking $\dot{M}\simeq 10^{16}$ 
gs$^{-1}$ as a typical mass accretion rate during outbursts, 
we can estimate  the maximum of the field strength from Eq.~(1)
\begin{equation}
B_{12, max}\simeq 0.15 (\frac{P_{\rm s}}{20\,{\rm
s}})^{21/16}\dot{M}_{16}^{15/16}M_1^{1/4}R_6^{-3}.
\end{equation}
Alternatively, Eq.~(2) presents another upper limit of the field strength with
$L_{\rm X}\sim 10^{36}$ ergs$^{-1}$,
\begin{equation}
B_{12, max}\simeq 1.2\times 10^{-2} L_{{\rm X},36}^4M_1^{-2}R_6^{6},
\end{equation}
where $L_{{\rm X},36}=L_{\rm X}/10^{36}$ ergs$^{-1}$. However, since $L_{\rm cr}$
is quite insensitive to $B$ (see Eq.~[2]), this may lead to large uncertainty in estimating $B$
from the observed $L_{\rm X}$. So in the following we will not consider
the result from Eq.~(8).

Currently there are about 10 confirmed SFXTs,
among which only 4 have spin periods detected \citep[][and references therein]{Sid11}. 
In Table 1 we list their orbital and spin periods ($P_{\rm orb}$ and $P_{\rm s}$). 
In column 4 we also present the their X-ray luminosities (in units of $10^{33}$ ergs$^{-1}$)
in the intermediate state, or the minimum X-ray luminosities, 
for which spectral information was available. The maximum 
field strengths, derived from Eqs.~(6) and (7), are shown in columns 5
and 6, respectively.  Here we relate
the X-ray luminosity with accretion rate via the formula $L_{X}=GM\dot{M}/R$, and
adopt the NS mass $M=1.4\,M_{\odot}$ and radius $R=10^6$ cm.

From Table 1 we find interesting distributions of the  NS magnetic fields.
IGR J11215-5952, IGR J16465-4507, and IGR J18483-0311 may have 
relatively high magnetic fields ($\sim$ a few $10^{11}-10^{12}$ G), 
comparable with those of typical HMXBs; while the magnetic field of AX J1841.0-0536
is considerably low ($\la 10^{10}$ G).
 
We notice that IGR J11215-5952 is actually a peculiar SFXT, since it is the only one 
so far showing strictly periodic outbursts \citep{Sid07}. 
Both IGR J16465-4507 and IGR J18483-0311 are classified as ``intermediate" SFXTs
\citep[][see also La Parola et al. 2010; Clark et al. 2010]{WZ07,Rom10b}, 
since the variations of their X-ray fluxes are significantly lower than that observed
 from the SFXT prototypes ($\sim 10^4-10^5$).  Hence it seems that 
 only AX J1841.0-0536 is a ``proper" SFXT. Besides this source,
lines of possible evidence for low-field NSs in SFXTs  
were also noted in \citet{Sg10} and \citet{Gr10}. 

%Since quite a few other SFXTs show similar X-ray luminosities around 
%$10^{33}-10^{34}$ ergs$^{-1}$, they are expected to have similar
%magnetic fields if their spin periods resides in tens to hundreds of 
%seconds. 

(3) Another possible estimate of the magnetic fields can be obtained
from the size $S$ of the polar caps during accretion \citep{FKR02},
\begin{equation}
S\simeq\pi R_{pc}^2=\pi(R/R_m)R^2\sin^2\alpha,
\end{equation}
where $R_{pc}$ and $\alpha$ are the radius of the polar cap, and the angle
between the magnetic axis and the equatorial plane, respectively. Eq.~(9) 
yields the maximum of the magnetospheric radius if $R_{pc}$ is known,
\begin{equation}
R_{m}\la (R/R_{pc})^2R.
\end{equation}
Combining Eqs.~(3) and (10) and taking $R_{pc}\sim 0.25$ km, 
$\dot{M}\sim 10^{13}$ gs$^{-1}$ 
\citep[e.g.][]{Rom10b}, 
we always have $B_{12}\la 0.1$, which seems to be independent on the models
of accretion modes of NSs. However, one should take caution when adopting
this result, since other spectral models (e.g., absorbed power-law, 
sometimes with high-energy cutoff) also well fits the data during the intermediate
and low states. Additionally,
\citet[][see also Bodaghee, et al. 2010]{Boz10} pointed out that, 
models describing spherical accretion onto magnetized NSs predict 
an inverse relation between the size of the hot spot and $L_{\rm X}^{1/2}$,
at least for systems with $L_{\rm X}<10^{35}$ ergs$^{-1}$ \citep{WSH83}.
This seems to be in contrast with the observed relation that the blackbody radius grows 
with the X-ray intensity \citep{Rom09a}.

The above estimates suggest that the NSs in some 
SFXTs could have relatively low fields ($\la$ a few $10^{10}-10^{11}$ G). 
This distinct feature may differentiate them from normal
supergiant HMXBs. Because of their low magnetic fields, they are able
to accrete at very low rate, resulting in a large dynamic range of X-ray 
luminosities.  We notice that recent {\em XMM-Newton} observations of 
IGR J18483$-$0311 showed evidence of an emission line feature at $\sim 3.3$ keV 
in the 0.5$-$10 keV spectrum, implying an NS magnetic field of 
$\sim3\times10^{11}$ G if it could be ascribed to an electron cyclotron emission line
\citep{Sg11}. However, We need to emphasize that in deriving both Eqs.~(6) and (7)
we have assumed a dipolar magnetic field for the NSs, neglecting possible multipolar
structures. Additionally, deviation of the NS radius from 1$0^6$ cm
may considerably influence the estimate of the field stregths. 

\section{Constraints on the initial spin periods}
If part of the SFXTs are relatively low-field objects, it is interesting to
see how they have evolved to the current spin periods. 
NSs in binaries are traditionally thought to be born as 
rapidly rotating ($\sim 10$ ms) radio pulsars, spinning down by magnetic dipole radiation
until the ram pressure of the ambient wind material from the companion star
overcomes the pulsar's radiation 
pressure at the gravitational radius. In the subsequent evolution the
NS will rapidly spin down by losing energy into the surrounding
envelope material, until either $P_{\rm eq}$ or $P_{\rm br}$ is reached,
and a bright X-ray source emerges \citep{DP81}. Note 
that most of the evolution should be finished when the companion star
is still unevolved. For a typical mass-loss rate of 
$10^{-8}-10^{-7} M_{\odot}$yr$^{-1}$ from a 30 $M_{\odot}$ main-sequence star 
and a wind velocity of  $v_{\rm w}\sim 10^3$ kms$^{-1}$, 
the accretion rate of the NS is $\sim 10^{15}-10^{16}$
gs$^{-1}$ if $P_{\rm orb}\sim 10-100$ d. Since the spin-down time during the 
radio pulsar phase occupies the majority of the spin-down evolution (Davies
\& Pringle 1981), we can
use it to estimate the total spin-down time,
\begin{equation}
\tau\simeq 8\times 10^{7}(B_{12}/0.1)^{-1}\dot{M}_{16}^{-1/2}v_8^{-1}\,{\rm yr},
\end{equation}
where $v_8=v_{\rm w}/10^8$ cms$^{-1}$. This is at least one order of magnitude
larger than the main-sequence lifetime ($\la 5\times 10^6$ yr) of the 
companion star, indicating that the current NSs should be in the pulsar
phase with periods less than $\sim 1$ s, obviously in contrast with
the observations of SFXTs. Hence the traditional spin-down evolution
mechanisms do not apply at least for some SFXTs.

The problem may be resolved in two kinds of ways. The first is that the
wind of the companion star was not  
spherically expending, but has had an asymmetric form, e.g., a disk-like
wind with high density and low velocity as observed in Be stars. In this respect, 
it is interesting to see that IGR J11215$-$5952 is indeed located in the Be/X-ray
binary region in the $P_{\rm orb}-P_{\rm s}$ diagram, and shows
periodic outbursts. 

The alternative (and probably more general) possibility is that the NS was born
rotating slowly ($P_{\rm s}\la 1$ s), so that it went directly into the propeller
phase after its birth. Propeller spin-down to either $P_{\rm eq}$ or $P_{\rm br}$ would
take a time less than the main-sequence lifetime of the massive
companion star \citep[][and references therein]{Li99},
\begin{equation}
\tau\simeq 2.7\times 10^{5}(B_{12}/0.1)^{-1/2}\dot{M}_{16}^{-3/4}P_1^{-3/4}\,{\rm yr},
\end{equation}
where $P_1=P_{\rm s}/1$ s.

In the latter case, these SFXTs will be distinct by relatively low fields and long initial
spin periods from normal young NSs. These  objects have been
so called ``anti-magnetars" (young NSs born with a weak 
dipole field), originally discovered in the compact
central objects (CCOs) in supernova remnants. For example,
1E 1207.4$-$5209, the peculiar CCO in the
supernova remnant G296.5+10.0, has been proposed to be an
anti-magnetar. Its spin period of 0.424 s \citep{Za00}
and the upper limit to its period derivative $\dot{P}< 2.5 \times
10^{-16}$ ss$^{-1}$ \citep{GH07} yield a dipole magnetic field $B< 3.3\times 10^{11}$ 
G. The characteristic age of the NS $\tau_{c}> 27$ Myr exceeds by 3 orders of 
magnitude the age of the supernova remnant, suggesting that 1E
1207.4$-$5209 was born with a
spin period very similar to the current one. Evidence for similar
low magnetic fields has been obtained for two other members of the
CCO class, namely CXOU J185238.6+004020 at the center of the Kes 79
SNR \citep{H07} and RX J0822-4300 in Puppis A \citep{GH09}.

We notice that the above arguments still suffer statistical problem, because
of very few SFXTs with magnetic field estimated. Indeed SFXTs do exhibit 
diverse behaviors. For all of the most extreme SFXTs (e.g., IGRJ17544$-$2619, 
IGRJ08408$-$4503, and XTEJ1739$-$302), the spin periods have not yet been 
detected, and in a few cases very long spin periods have been suggested 
(up to $1000-2000$ s, see Smith et al. 2006) and Eqs.~(6) and (7) suggest
that the magnetic field strengths can be as high as a few $10^{13}$ G 
\citep[see also][]{Boz08}. For those SFXTs that host very slowly spinning NSs, 
there is no problem to spin down to current periods within $\sim 10^7$ yr 
\citep{Zh04}. If confirmed, they indicate that SFXTs may belong to different classes of 
objects and their origins will be an interesting subject for future investigation.

\section{Conclusions}

We have derived possible constraints on the upper limit of magnetic
field strengths of NSs in 4 SFXTs from recent observational results.
Among them the only ``real" SFXT AX J1841.0$-$0536 may have relatively low field
($\la 10^{10}$ G) and long initial spin period ($\sim 1$ s), while the field strengths of 
the the other three are considerably 
stronger. If  AX J1841.0$-$0536 is not an extreme case, it suggests that the NSs in some SFXTs 
may represent another kind of anti-magnetars besides the CCOs discovered 
in supernova remnants, and  have interesting implications for the formation of
 young NSs born with relatively low fields and long spin periods.
Future observations, especially
the detection of the cyclotron features in the X-ray spectra of SFXTs and monitoring their 
spin evolution to constrain the accretion torques will be crucial in unveiling the nature 
of the NSs in SFXTs.

\section{Acknowledgements}
%We are grateful to an anonymous referee for helpful comments and suggestions. 
This work was supported by the Natural Science Foundation of China
(under grant number 10873008) and the National Basic Research
Program of China (973 Program 2009CB824800).

%\begin{figure}
%\epsscale{0.8}
%\plotone{V1P01.eps}
%\caption{Calculation down by standard parameters of $V_w=1000km/s$ and $P_s=0.1s$}
%\end{figure}

%\begin{figure}
%\epsscale{0.8}
%\plotone{changes.eps}
%\caption{the upper two are done by enlarged $P_s$ of $P_s=1.0s$ and $P_s=5.0s$, while the bottom two are got from $V_w=1500km/s,2000km/s$ with unchanged $P_s$.}
%\end{figure}
\newpage
\begin{table}
\caption{Observed and derived parameters of 4 (possible) SFXTs}
\begin{tabular}{cccccccc}
\hline\hline
Source     &  $P_{\rm orb}$ (d)  & $P_{\rm s}$ (s) & $L_{\rm X,33}$ &
$B_{12, max}$ & $B_{12, max}$ &  Ref \\
 \hline
IGR J11215-5952   &  165     & 186.8   & $\sim10$ & 3.5 & 2.9  & 1
\\
IGR J16465-4507  &  30.3 &  $\sim 228$   & $\sim40$ & 8.9 & 3.73  & 2
\\
AX J1841.0-0536   &    &    4.74 & $\sim 2$ &  $2.2\times 10^{-2}$ & $2.3\times 10^{-2}$ & 3
\\
IGR J18483-0311   &  18.55   & 21.05 &$\sim10$ & 0.28 &  0.17 &  4
\\ \hline
\end{tabular}\\

References. 1. Romano et al. (2009b); 2. Zurita Heras \& Walter (2004); 
3. Romano et al. (2009a); 4. Giunta et al. (2009).
\end{table}

\end{document}